\documentclass[aps,prx,twocolumn,showpacs,citeautoscript,superscriptaddress,longbibliography,nofootinbib]{revtex4-1}
\usepackage{braket}
\usepackage{graphicx}  
\usepackage{epstopdf}
\usepackage{dcolumn}   
\usepackage{bm}        
\usepackage{amsmath}
\usepackage{amssymb}   
\usepackage{wrapfig}
\usepackage{array}
\usepackage{comment}
\usepackage{bbding}

\usepackage{hyperref}
\begin{document}
\title{
  Chern-Simons type cross-correlations and geometric Born effective charge of phonons
}

\author{Swati Chaudhary}\email{swatichaudhary@issp.u-tokyo.ac.jp}
\affiliation{The Institute for Solid State Physics, The University of Tokyo, Kashiwa, Chiba 277-8581, Japan}


\author{Takashi Oka}
\email{oka@issp.u-tokyo.ac.jp}
\affiliation{The Institute for Solid State Physics, The University of Tokyo, Kashiwa, Chiba 277-8581, Japan}

\begin{abstract}

The interplay between different degrees of freedom in condensed matter systems engenders a rich variety of emergent phenomena. In particular, fermions with non-trivial quantum geometry  can generate  Chern-Simons (CS)-like terms in effective field theories for different gauge fields. For phonons, such terms can result in chiral phonon splitting. 
Here, we propose that the local Berry curvature can influence the spectra and dynamics of optical phonons, even in materials with zero Chern number, which we demonstrate with a gapped Dirac model. 
We identify a previously overlooked CS like cross-correlation between electromagnetic and pseudo-gauge fields in 2+1 dimensions which depends on valley Chern number. It facilitates a direct coupling between phonons and photons by inducing a geometric Born effective charge.  This opens up a new route for coherent Raman phonon excitation and quantum geometry probes.

\end{abstract}
\maketitle

\section{Introduction}
The interaction between electrons and the crystal lattice underlies many important phenomena in condensed matter physics, including superconductivity, polaron formation, and charge density waves. Phonons can profoundly influence electronic transport and optical responses~\cite{ziman2001electrons}. Conversely, electronic degrees of freedom can significantly modify phonon spectra, their response to external perturbations, and even give rise to new routes for coherent phonon excitation~\cite{Kohn1959,Zeiger1992,dhar1994time,Ron2020}.  When electrons are integrated out, electron-phonon coupling can manifest as entirely new terms in the effective phonon action, imprinting signatures of broken symmetries from the electronic or spin sector onto the phonons~\cite{Liu2021,Che2025PRL,wu2025magnetic,Ren2021,ZhangPRL2023,sutcliffe2025pseudo}.

It is well established that the quantum geometric properties of electronic bands can give rise to a wide range of unconventional electromagnetic phenomena~\cite{XiaoRev2010,Fukushima2008} such as anomalous Hall~\cite{HaldanePRL1988}, different quantum anomalies~\cite{nielsen1983adler,Niemi1983,Son2013,burkov2015chiral,Tanwar2023}, and  axion electrodynamics~\cite{Wilczek1987,Essin2009}.  These effects are very elegantly captured by additional terms in low-energy effective field theories for electromagnetic fields~\cite{Qi2008}, resulting from the integration of fermionic degrees of freedom. Many of these phenomena, often, appear in the form of emergent Chern-Simons (CS)-like couplings, and are directly linked to first or second-Chern number of electronic bands~\cite{Taherinejad2015,Nomura2012}.

More recently, similar geometric contributions have been recognized to be crucial for phonons and their responses to external perturbations. In particular, electronic Berry curvature can generate new ``kineo-elastic" terms in the effective phonon action~\cite{dong2025phonons,Liu2021,Ren2021,ZhangPRL2023,Barkeshli2012}. In certain systems with finite Chern number, these terms manifest as phonon-Hall viscosity resulting in non-degenerate chiral phonons and giant phono-magnetic responses~\cite{Chen2025chiral,cheng2020large}.  Such Chern-Simons like terms in effective phonon action can arise in systems such as graphene, transition-metal dichalcogenides (TMDs), different Dirac or Weyl semi-metals, where phonons behave like pseudo-gauge fields.  Based on this analogy with electromagnetic fields, previous works have identified a direct relation between electronic DC Hall conductivity and phonon-Hall viscosity~\cite{Chen2025chiral,Liu2017WSM,Barkeshli2012}, which links phonon chirality with Chern number of electronic bands. 

This naturally raises some important questions: \textit{Can local quantum geometric features of electronic bands leave observable signatures on optical phonons, even in systems with zero net Chern number? Are there any important consequences of local Berry curvature for phonon dynamics?} In this work, we explore these questions by considering the coupling of gapped Dirac fermions with optical phonons which behave like pseudo-gauge fields. 

We first show that the optical phonons with finite frequency can sense local Berry curvature, which leads to a splitting of chiral phonons in a honeycomb lattice where inversion symmetry breaking is stronger than the time-reversal symmetry breaking. Unlike previous works which considered zero frequency limit, this effect arises when the frequency dependence of the emergent CS contribution from the one-loop diagram of gapped Dirac fermions is carefully taken into account. Although, it still requires broken TRS,  the effect can persist in systems with zero Chern number. It leads to phonon chirality and splitting  which offers a potential route to probe the Floquet generated Haldane mass with phonons at modest drive strengths or to detect weaker TRS breaking effects in electronic systems.

Most importantly, we introduce a previously unrecognized Chern-Simons-type cross-coupling between electromagnetic fields and pseudo-gauge fields generated by optical phonons, explicitly mediated by local Berry curvature. While previous works have investigated acoustoelectric phenomena originating from similar coupling mechanism which leads to current generation from  acoustic waves~\cite{Bhalla2022,PhysRevLett.134.026304} or time-dependent strain~\cite{Vaezi2013}, and piezoelectric effects~\cite{Droth2016,rostami2018piezoelectricity}, we focus on the direct coupling between photons and optical phonons induced by local Berry curvature. This mechanism imparts a geometric Born effective charge (BEC) to an otherwise Raman-like phonon. It remains present even in systems with vanishing total Chern number, and is directly related to the Valley Chern number. We analyze the consequences of this coupling for phonon dynamics and show that it enables a new mechanism for direct excitation of coherent Raman modes in graphene. 

Next, we study the influence of this geometric BEC in a gapped honeycomb model with an additional weak TRS breaking term and demonstrate that the driven phonon mode can acquire non-zero chirality even for a linearly polarized THz drive, circumventing the need for circular polarization optics. Interestingly, the sign of chirality in this case is controlled by the detuning between natural phonon frequency and the frequency of THz drive.  We believe this type of cross-coupling may be present across a wide variety of materials and can involve orbitals and spin degrees of freedom as well, given the ubiquity of Chern-Simons-like couplings in diverse systems~\cite{Nomura2010,Garate2010}.

\section{Dirac system, gauge fields, and pseudo-gauge fields:}
We consider the low-energy Dirac fermions ($\psi$ ) around $K$ and $K'$ valleys in graphene coupled to non-dispersive $E$ mode optical phonons.  The phonon coordinates $(Q_x,Q_y) $ are the dynamical degrees of freedom, also represented by pseudo-gauge fields $(a_x,a_y)=g/ev_F\,(Q_y,-Q_x) $ in 2+1 dimensions~\cite{vozmediano2010gauge}. We also add the $U(1)$ gauge fields $A_\mu$ for electromagnetic fields coupled to the electron. The system can be described as 
\begin{align}
\mathcal{L} &= \mathcal{L}_{\text{Dirac}} + \mathcal{L}_{EM} + \mathcal{L}_{\text{ph}} + \mathcal{L}_{\text{int}} \nonumber\\&= \bar{\psi} \left( \gamma^0 p_0 - v_F \bm{\gamma} \cdot \bm{p} - M \right) \psi \}-\frac{1}{4}F_{\mu\nu}F^{\mu\nu} \nonumber \\
&\quad +  \sum_{\mu=1,2} \frac{e^2v_F^2}{g^2}\frac{\rho_I}{2} \left( \left( \partial_0 a_\mu \right)^2 - \omega_0^2 a_\mu ^2 \right)\nonumber \\
&\quad  +ev_F\bar{\psi} \gamma^\alpha \gamma^5 \psi a_\alpha \nonumber +ev_F  \bar{\psi} \gamma^\alpha \psi A_\alpha
\end{align}
where $\alpha = 1,2$ denotes spatial indices and $\gamma^\alpha p_\alpha = -\bm{\gamma} \cdot \bm{p}$, $\bar{\psi} = \psi^\dagger \gamma^0$, $\rho_I$ is the ion mass density in graphene, $v_F$ is the Fermi velocity, $g$ is the electron-phonon coupling, and $\omega_0$ is phonon frequency.  The gap matrix  $M = m_D + m_{AB} \gamma^3 + m_H \gamma^3 \gamma^5$,     where in graphene, $m_{AB}$ denotes the A, B sublattice potential (Semenoff mass), $m_H$ is the Haldane mass which can be induced by an external drive, and $m_D$ is the Dirac mass which mixes the two valleys and has been taken to be zero here.

We begin by integrating out Dirac electrons which results in the following correction to the effective action for the gauge fields $A$ and $a$ in the Fourier space:
\begin{align}
    \mathcal{S}_{eff}^{(1)}(A,a)&=&\sum_{\alpha,\beta=A,a} \frac{e^2}{\hbar v_F^2}\,\int \frac{d^3 p}{(2\pi)^3}\left[\alpha^\mu(-p)\Gamma^{\mu\nu}_{\alpha\beta}(p)\beta^\nu (p)\right]
\end{align}
where $p=(\omega,v_F\bm{p})$, and $\mu,\rho,\nu=0,1,2$. This  emergent coupling 
takes Chern-Simons form, $\Gamma^{\mu\nu}_{\alpha\beta}(p)=\epsilon^{\mu\nu\rho}p^\rho\Pi^{\mu\nu}_{AA}(p)$ where
\begin{eqnarray}
  \Pi^{\mu\nu}_{AA}(p)=\sum_{i=1,2}\frac{m_i}{2\pi |p|} \text{arcsinh}\left(\frac{|p|}{\sqrt{p^2+4m_i^2}}\right)\\
  \Pi^{\mu\nu}_{Aa}(p)=\sum_{i=1,2}\frac{\tilde{m}_i}{2\pi |p|} \text{arcsinh}\left(\frac{|p|}{\sqrt{p^2+4\tilde{m}_i^2}}\right)
  \label{Eq:CSphotonphonon}
\end{eqnarray}
with $p^2=\omega^2-v_F^2|\bm{p}|^2$, $ \Pi^{\mu\nu}_{aa}(p)= -\Pi^{\mu\nu}_{AA}(p)$, $ \Pi^{\mu\nu}_{aA}(p)= \Pi^{\mu\nu}_{Aa}(p)$ and  $m_{1(2)}=m_H\pm m_{AB}$, $\tilde{m}_{1(2)}=m_{AB}\pm m_H$ (check Appendix for derivation).
In the long wavelength ($p\rightarrow 0$) limit,
\begin{align}
    \Gamma^{\mu\nu}_{AA}(p)\approx\epsilon^{\mu\nu\rho}p^\rho\frac{1}{2\pi} \sum_{i=1,2}\left(c(m_i)+p^2\,d(m_i)+O(p^4)\right)\\
    \Gamma^{\mu\nu}_{Aa}(p)\approx\epsilon^{\mu\nu\rho}p^\rho\frac{1}{2\pi} \sum_{i=1,2}\left(c(\tilde{m}_i)+p^2\,d(\tilde{m}_i)+O(p^4)\right)
    \label{Eq:Gammamunu}
\end{align}
where $c(m)=\text{sign}{(m)}/2$ and $d(m)=-\text{sign}{(m)}\hbar^2/(12\,m^2)$. 
The CS term for photon-photon and phonon-phonon field is dictated by Haldane mass and is proportional to total Chern number $C=c(m_1)+c(m_2)$ which is a well known contribution in QED and have been identified as crucial for phonons as well~\cite{Chen2025chiral,Barkeshli2012}. On the other hand, the Semenoff mass leads to a CS like cross-correlation between photon and phonon fields which depends on valley Chern number, $\tilde{C}=c(\tilde{m}_1)+c(\tilde{m}_2)$, instead. 
Additionally, we will soon see that the quadratic term $d(m)$ also plays a crucial role when the total Chern number $C = 0$. In the following sections, we discuss the consequences of these different terms on phonon chirality, spectra, and their dynamics under an external electromagnetic field.

\begin{figure}
    \centering
       \includegraphics[scale=0.65]{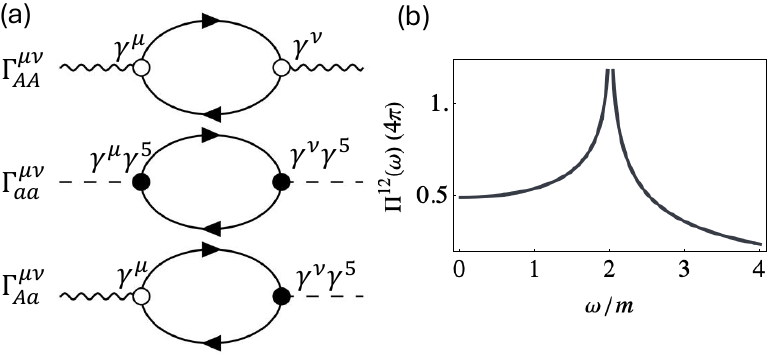}

    \caption{(a) One-loop diagram responsible for  photon-photon, phonon-phonon, and photon-phonon  Chern-Simons coupling, representing the interaction between different gauge fields (b) Contribution $\Pi^{12}$ as a function of gauge field frequency $\omega$. It remains quantized for frequency much smaller than the Dirac gap $2m$ and shows a resonance like behavior at $\omega=2m$. We avoid the resonant and high-frequency scenario in our calculations but even at small frequencies, there is a weak quadratic frequency dependence which is relevant to observe a chiral phonon splitting in globally trivial phase.}
    \label{fig:Feynman}
\end{figure}

\subsection{Spectrum of chiral phonons}
We begin by investigating the implications of phonon-phonon CS term on phonon spectra. The bare Green's function for $E$ mode optical phonons is $D(\omega)=\text{Diag}[D_x (\omega),D_y(\omega)]$, where $D_x(\omega)=D_y(\omega)=2\omega_0/(\omega^2-\omega_0^2)$. 
This CS like emergent coupling leads to an off-diagonal self-energy, $\Sigma_{xy}(\omega)=-\Sigma_{yx}(\omega)=i 2\omega g^2/(2\omega_0 v_F^2\rho_I) \Pi^{12}_{aa}(\omega)\,\,$ which is represented by Feynman diagrams in Fig.~\ref{fig:Feynman} (a). This term splits the degeneracy and the resulting modes are chiral in nature. In monolayer graphene,  $g\approx10$ eV/$\AA$, $v_F\approx10^6$ m/s, and mass density  \( \rho_I \approx 7.6 \times 10^{-7} \, \mathrm{kg/m^2} \) which gives $2 g^2/(2 \pi v_F^2\rho_I)\approx 0.7 \,\rm{ meV} \ll \omega_0$. This allows us to approximate the splitting $\Delta \omega \approx 2 g^2/( v_F^2\rho_I) \Pi^{12}_{aa}(\omega)$ whose frequency dependence is shown in Fig.~\ref{fig:Feynman} (b). 
In the limit of $\omega\ll |m_{AB}\pm m_H|$,  $\Pi^{12}_{aa}(\omega)$ term is quantized to $C/2\pi$ where $C$ is the Chern-number, and the splitting $\Delta\omega\approx 0.7\,\rm{meV}$ in Haldane phase which is equal to the saturated splitting obtained from external magnetic field in Ref.~\cite{Chen2025chiral}. 

Similar results can also be reached by considering the equation of motion for phonons. For that purpose, first we return to the coordinate space where $p=(-i\partial_t,-iv_F\nabla)$, and keep only first few orders of derivatives. Since, we are focusing on Gamma phonons, so the derivatives of the fields with respect to position are also ignored (check appendix for these spatially varying terms).  After including the corrections from CS term, phonon Lagrangian becomes, $\mathcal{L}_{ph}=\mathcal{L}^{0}_{ph}+\mathcal{L}^{CS}_{ph}$ with
\begin{equation}
    \mathcal{L}^{CS}_{ph}=\frac{g^2}{hv_F^2}\,\left[C (Q_x\dot{Q}_y-Q_y\dot{Q}_x)+D (Q_x\dddot{Q}_y-Q_y\dddot{Q}_x)\right],
\end{equation}
where CS term appears as phonon-Hall viscosity, and in the chiral basis, $Q_{L/R}=(Q_x\pm iQ_y)/\sqrt{2}$, the equation of motion becomes (check Appendix D):
\begin{equation}
    \ddot{Q}_{L/R}+\omega_0^2 Q_{L/R}=\pm 2 i \frac{g^2}{hv_F^2\rho_I}\left(  C \dot{Q}_{L/R}+ D\dddot{Q}_{L/R}\right), \label{Eq:EOM2}
\end{equation}  resulting in a frequency splitting
\begin{equation}
  \hbar(  \omega_L-\omega_R)\approx 2\frac{g^2}{2\pi v_F^2\rho_I}\left(C-D\omega_0^2\right).
\end{equation}
The first contribution is associated with the Chern number and was identified in previous works~\cite{Chen2025chiral} which established a direct link between DC Hall conductivity and phonon Hall viscosity leading to chiral phonon splitting. 

\begin{figure}
    \centering
     \includegraphics[scale=0.5]{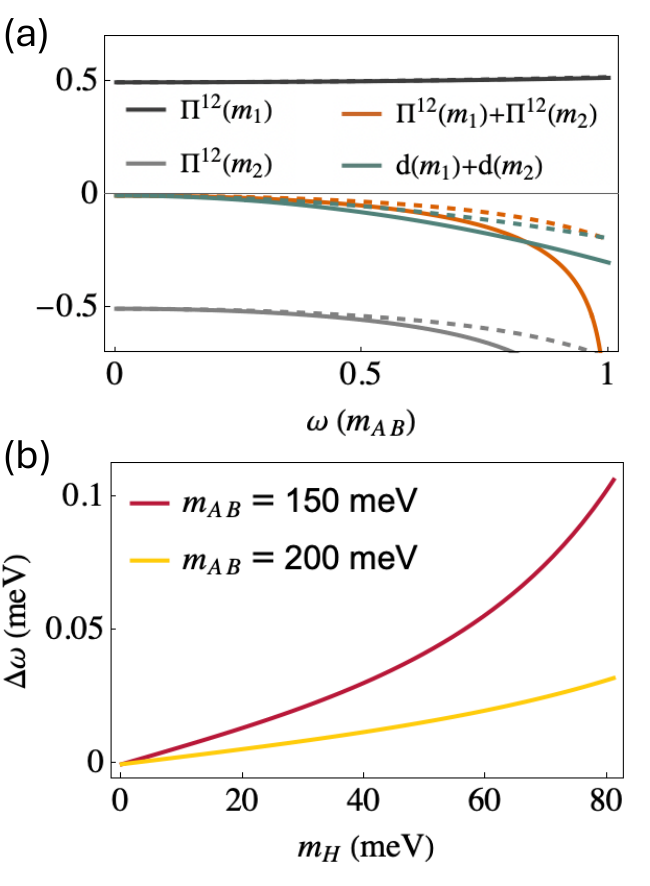}

    \caption{(a) Contributions to $\Pi^{12}$ from the two Dirac points of graphene, with masses $m_{1(2)} = m_{AB} \pm m_H$, are shown individually, along with their sum and a quadratic-frequency approximation for $m_H=0.5\, m_{AB}$ (solid) and $m_H=0.4\, m_{AB}$ (dashed). (b) Chiral phonon splitting of $E$ mode ($\omega_0=100$ meV) as a function of Haldane mass for two different Semenoff masses. The splitting increases monotonically as $m_H$ increases and the effect is more pronounced for smaller $m_{AB}$. Parameters are chosen such that phonon frequency, $\omega_0 <2|m_{AB}\pm m_H|$ and is off-resonant with both  gaps.}
    \label{fig:splittingandPi}
\end{figure}
When the finite frequency of the phonons is taken into account, the term $\Pi^{12}_{aa}(\omega)$ is no longer quantized (Fig.~\ref{fig:Feynman}~(b)) and the deviation can be approximated by a secondary contribution denoted by $D$ which depends on the magnitude of the mass term unlike Chern number as shown in Fig.~\ref{fig:splittingandPi} (a). Interestingly, this contribution can survive even when the total Chern number is zero, provided that TRS is broken and the two gaps \( |m_H \pm m_{AB}| \) are unequal, which is usually the case for the AC Hall conductivity as well.   The Chern number contribution is dominant when Haldane gap \( 2m_H \gg \omega_0 \approx 100 \, \mathrm{meV} \), which is difficult to realize in graphene or TMDs. Since the analysis assumes \( \omega_0 \ll 2|m_{AB}\pm m_H|\), the second term may be relevant in systems such as graphene on hBN or TMDs under circularly polarized light, where a Haldane mass of the order of $10-100$ meV can be generated~\cite{mciver2020light}.

Considering, \( m_{AB} \approx 100- 500 \, \mathrm{meV} \), phonon frequency \( \omega_0 \approx 100 \, \mathrm{meV} \), and  approximating other parameters with that of monolayer graphene, a light-induced Haldane mass could lead to a chiral phonon splitting as shown in Fig.~\ref{fig:splittingandPi} (b). Depending on the relative values of $m_H$ and $m_{AB}$, a splitting of the order of $0.1$ meV can be achieved at $m_H\approx 80$ meV, whereas the global topology remains trivial.  In contrast, for Haldane phase with \( C = 1 \) and \( 2m_H \gg \omega_0 \), a maximum splitting of \( \Delta \omega \approx 0.7 \, \mathrm{meV} \) was calculated.  This analysis shows that TRS breaking in electronic sector can induce phonon chirality even though the electronic system remains topologically trivial. Many recent works have demonstrated that the ability of chiral phonons to inherit signatures of time-reversal symmetry breaking is responsible for their large phononic magnetic moment and may serve as a probe for identifying different types of time-reversal symmetry breaking orders in the system~\cite{sutcliffe2025pseudo,Chaudhary2024,lujan2024spin,Liu2021,Che2025PRL,wu2025magnetic,hernandez2022chiral,mustafa2025origin}.

\subsection{Dynamics and excitation mechanism}
Next, we investigate the consequences of photon-phonon coupling and study the dynamics of phonons driven by AC light. 
We established in Eq.~\ref{Eq:CSphotonphonon}, that a Chern-Simons like coupling can emerge between photons and phonons for gapped Dirac fermions. In presence of this cross-correlation, the equations of motion for phonon displacement supplemented with the damping term are given by (check Appendix~\ref{appendix:EOM}):
\begin{eqnarray}
\begin{split}
 &\ \ddot{Q}_{L/R} + \eta\dot{Q}_{L/R}+\omega_0^2 Q_{L/R} =\\& \mp i\, \frac{2 g^2}{2\pi\hbar v_F^2 \rho_I}\left( C \dot{Q}_{L/R} + D\dddot{Q}_{L/R} \right) \\&+\,\frac{2\,eg}{2\pi\hbar v_F\rho_I}\left(\tilde{C}\dot{A}_{L/R}+\tilde{D}\dddot{A}_{L/R}\right).
\end{split}
\end{eqnarray}
where $\eta$ denotes the damping coefficient and $\hbar\eta\sim 0.1-1$ meV for TMDs and multilayer graphene. This indicate that a non-zero valley-Chern number, $\tilde{C}$ assigns a non-zero BEC to the phonon leading to a direct linear coupling with laser fields. For graphene, with unit cell area $\mathcal{A}_{\text{unit cell}}\approx 5.25\, \AA^2$, this geometric BEC, $Z^*= 2eg/(2\pi\hbar v_F)\times \mathcal{A}_{\text{unit cell}}\approx 2.5\, e$ which is comparable to BEC of infrared phonons in hBN~\cite{guo2007static}.  We investigate different scenarios which can arise in graphene/TMDs or Chern-insulators.\\

\begin{figure}
    \centering
     \includegraphics[scale=0.64]{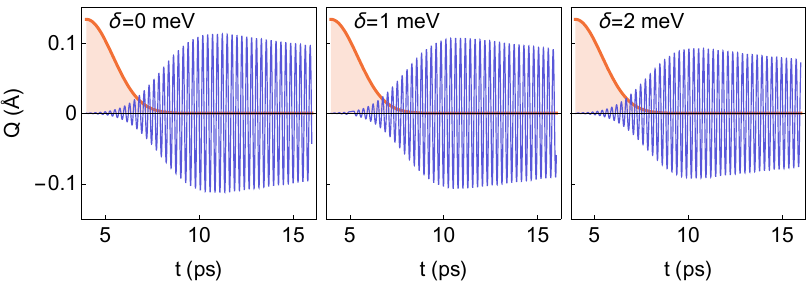}
     \caption{Phonon displacement at three different detuning frequencies for peak amplitude $E=1$ MV/cm of laser pulse shown by red envelope, phonon damping $\eta= 0.3$ meV, and $m_H=0$. This phonon dynamics is induced by local Berry curvature which mediates coupling between light and Raman-like phonons in graphene with finite Semenoff mass.}
     \label{fig:phonondynamics}
\end{figure}

\textit{Dynamics in the absence of Haldane mass:} In most common scenarios, graphene or TMDs do not break TRS and $m_H=0$. In this case,  the phonon-phonon CS coupling (or Hall-viscosity terms) terms vanish, i.e., $C=D=0$. However, a direct CS like coupling between photons and phonons can be mediated  by local Berry curvature. This can lead to a new excitation mechanism for optical phonons in graphene which do not carry any Born effective charge, and thus do not couple directly to light fields. The resulting dynamics is shown in Fig.~\ref{fig:phonondynamics} indicating a lattice vibration amplitude of the order of $0.1\, \AA$ for a laser pulse of 2 ps and a maximum electric field, $E= 1\, \rm{MV/cm}$. The phonon polarization is same as the polarization of EM field, and the dynamics is identical for right- and left -circularly polarized lattice vibrations at any given frequency.  

\begin{figure}
    \centering
     \includegraphics[scale=0.67]{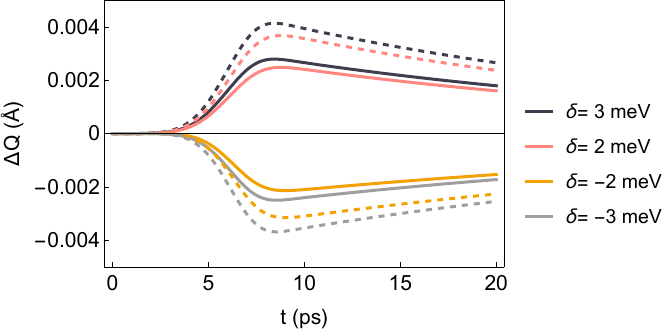}
     \caption{Difference in phonon displacement amplitudes between two chiralities when driven by linearly polarized light for $E=1$ MV/cm, $\eta= 0.3$ meV and $m_H=80$ meV (solid) and $m_H= 90$ meV (dashed). We have considered $m_{AB}=150 $ meV.}
     \label{fig:dichroism}
\end{figure}

\noindent{\textit{Non-zero Haldane and Semenoff mass:}}
Here, we consider another scenario, where the Chern-number is zero but Haldane mass is non-zero. 
In this case, the phonon energies would split due to the second-order corrections (denoted by $D$ in Eq.~\ref{Eq:EOM2}), and if we take a linearly polarized light, we can expect a slight difference in phonon amplitudes for two chiralities. Furthermore, this difference can be exploited to get a small non-zero net chirality whose sign can be controlled by the sign of detuning. In Fig.~\ref{fig:dichroism}, we show the different between the amplitude of right- and left-circularly polarized lattice vibrations when the phonons are driven by a linearly polarized light with frequency $\omega=\omega_0+\delta$. The difference in amplitudes is not that large but it can facilitate the study of chiral phonons with linearly polarized light. This small difference can perhaps be captured in transient-reflectivity measurement in a pump-probe spectroscopy.

\section{Discussion}
In this work, we establish that local Berry curvature can not only induce chiral phonon splitting but can also open up a new route to couple phonons with photons even in systems with zero Chern number. This coupling is very elegantly captured by Chern-Simons like cross-correlations and can arise in gapped Dirac systems. It serves a dual purpose, as it can generate a large BEC by engineering electronic quantum geometry and also provides a pathway to probe local quantum geometric features of electronic bands via phonon spectroscopy. The cross-correlations explored in this work are also expected to influence the dynamics of the electromagnetic field, potentially leading to linear or circular birefringence depending on the polarization of the excited phonon mode which we leave to the future work.

We anticipate that similar cross-correlations may exist between various types of gauge fields across a wide variety of materials~\cite{Cortijo2015, araki2023emergence}. Beyond graphene, pseudo-gauge fields can also arise due to spin textures, moire defects, additional degrees of freedom, or other spatial and temporal modulations~\cite{Ryu2012,TrifunovicPRB2019,Garaud2012,Araki2021}, potentially leading to direct coupling between electromagneic and spin or other degrees of freedom. Moreover, even in  graphene, modulation of spin-orbit coupling or charge density can lead to non-Abelian generalization of such gauge fields~\cite{Tokatly2008,Gopalakrishnan2012}. 
Weyl semimetals can be another possible paltform to study such cross-correlations where pseudo-gauge fields emerge naturally due to strain or magnetic texture, which act as axial gauge fields~\cite{Cortijo2015,ozawa2024chiral,araki2020magnetic}. These fields can drive anomalous transport and could modify phonon properties. Extending our framework to such systems could shed light on phonon-mediated signatures of chiral anomaly and open possibilities for optically accessing Weyl band topology through lattice degrees of freedom.

Lastly, it might be interesting to explore multilayer graphene systems which offer easy tunability of quantum geometric features and the layer degree of freedom can allow for non-abelian gauge fields~\cite{josePRL2012}. In these systems, the quantum metric of the electronic bands also plays an important role, contributing to both linear and nonlinear optical conductivities~\cite{nagaosa2017concept,ahn2022riemannian,Torma2023,Araki2018}. Similar effects can be expected for phonons, where the quantum geometry may influence phonon peaks and linewidths~\cite{Sasaki2018} and may lead to second- or higher-order phonon-photon processes. This field-theoretical framework can be readily extended to analyze such processes.

\section{Acknowledgments}
S.C and T.O. acknowledge support from JSPS KAKENHI (No. JP23H04865, No. 23K22487), MEXT, Japan.


\onecolumngrid
\appendix
\section{Graphene Dirac points and field theoretic model}
In order to illustrate the $\gamma$ matrix structure of Semenoff mass and Haldane term, first we rewrite the effective low-energy hamiltonian at $K/K'$ valley in a $4\times4$ notation. For a given valley denoted by $\chi=\pm$ for (K/K'), we have:
\begin{equation}
    H_{\chi}=m_{AB}\sigma_z+\chi m_{H}\sigma_z+v_Fk_x\sigma_x+\chi v_Fk_y\sigma_y \equiv \begin{pmatrix}
       m_{AB}\sigma_z+ m_{H}\sigma_z+v_Fk_x\sigma_x+ v_Fk_y\sigma_y &0\\0& m_{AB}\sigma_z- m_{H}\sigma_z+v_Fk_x\sigma_x- v_Fk_y\sigma_y 
    \end{pmatrix}
\end{equation}
Next, we can rotate the lower block around $y$ axis which gives:
\begin{equation}
    H_{Dirac}= \begin{pmatrix}
       m_{AB}\sigma_z+ m_{H}\sigma_z+v_Fk_x\sigma_x+ v_Fk_y\sigma_y &0\\0& -m_{AB}\sigma_z+m_{H}\sigma_z-v_Fk_x\sigma_x- v_Fk_y\sigma_y
    \end{pmatrix}=\tau_z\left(m_{AB}\sigma_z+\vec{k}\cdot\sigma\right)+m_{H}\tau_0\sigma_z
\end{equation}
and next rotating $\tau_z$ to $\tau_x$, we get:
\begin{equation}
H_{Dirac}=m_{AB}\tau_x\sigma_z+\tau_x v_F\vec{k}\cdot\sigma+m_{H}\tau_0\sigma_z.
    \end{equation}
    which can be written as:
\begin{equation}
H_{Dirac}=\gamma^0\left(m_{AB}\gamma^3+v_F\vec{k}\cdot\gamma+m_{H}\gamma^3\gamma^5\right).
    \end{equation}    
in terms of $\gamma$ matrices
\begin{equation}
\gamma^0=\tau_z;\,\,\gamma^i=i\tau_y\sigma_i;\,\,\gamma^5=i\gamma^0\gamma^1\gamma^2\gamma^3=\tau_x\sigma_0
\end{equation}
which gives 
\begin{equation}
   \gamma^0\gamma^i=\tau_x\sigma_i;\,\, \gamma^3\gamma^5=\tau_z\sigma_z;\,\,\gamma^0\gamma^3\gamma^5=\tau_0\sigma_z.
\end{equation}
These  $\gamma$ matrices satisfy the following relations
\begin{equation}
   \{\gamma^\mu,\gamma^\nu\}= \text{Tr}(\gamma^\mu\gamma^\nu)=2g^{\mu\nu};\,\,\text{Tr}(\gamma^\mu\gamma^\nu\gamma^\rho\gamma^\sigma\gamma^5)=-4i\epsilon^{\mu\nu\rho\sigma},\,\{\gamma^\mu,\gamma^5\}=0
\end{equation}
and also for the special case where
$\mu,\nu,\rho$ are constrained to be in 2+1 dimensions,  
\begin{equation}
    \text{Tr}(\gamma^\mu\gamma^\nu\gamma^\rho\gamma^3\gamma^5)=-4i\epsilon^{\mu\nu\rho},\,\,\,\text{and  } [\gamma^{\mu},\gamma^3\gamma^5]=0,\,\,\,\text{for   } \mu,\nu,\rho={0,1,2}.
\end{equation}

\section{Effective Lagrangian derivation}
\subsection{Photon-Photon terms}
The corrections to photon-photon part of Lagrangian are dictated by the following term:
\begin{equation}
    \Pi^{\mu\nu}_{AA}(p)=\text{Tr}\left(\gamma^\mu\frac{1}{\gamma^\rho(k+\frac{p}{2})_\rho-m_D-m_{AB}\gamma^3-m_H\gamma^3\gamma^5}\gamma^\nu\frac{1}{\gamma^\rho (k-\frac{p}{2})_\rho-m_D-m_{AB}\gamma^3-m_H\gamma^3\gamma^5}\right).
\end{equation}
and we can write
\begin{equation}
    \frac{1}{\gamma^\rho k_\rho-m_D-m_{AB}\gamma^3-m_H\gamma^3\gamma^5}=\left(\gamma^\rho k_\rho+m_D-m_{AB}\gamma^3+m_H\gamma^3\gamma^5\right)\frac{1}{k^2-M^2-2m_Dm_H\gamma^3\gamma^5+2m_{AB}m_H\gamma^5}
    \label{Eq:PropagatormD}
\end{equation}
where $k^2=k_0^2-k_x^2-k_y^2$, $M^2=m_D^2+m_{AB}^2+m_H^2$ and we have used $\{\gamma^\mu,\gamma^\nu\}=2g^{\mu\nu}$, $\{\gamma^3,\gamma^3\gamma^5\}=0$, and $[\gamma^\mu,\gamma^3\gamma^5]=0$ for $\mu=0,1,2$.

However, in this work, we focus on $m_D=0$, where
\begin{equation}
\begin{split}
    \frac{1}{\gamma^\rho k_\rho-m_{AB}\gamma^3-m_H\gamma^3\gamma^5}=\left(\gamma^\rho k_\rho-m_{AB}\gamma^3+m_H\gamma^3\gamma^5\right)\frac{1}{k^2-M^2+2m_{AB}m_H\gamma^5}\\=\frac{\left(\gamma^\rho k_\rho-m_{AB}\gamma^3+m_H\gamma^3\gamma^5\right)\left(k^2-M^2-2m_{AB}m_H\gamma^5
    \right)}{(k^2-M^2)^2-(2m_{AB}m_H)^2}\\
    =\frac{\gamma^\rho k_\rho (k^2-M^2)-m_{AB}(k^2-M^2+2m_H^2)\gamma^3+m_H(k^2-M^2+2m_{AB}^2)\gamma^3\gamma^5-2m_{AB}m_H\gamma^\rho k_\rho\gamma^5}{(k^2-(m_H-m_{AB})^2)(k^2-(m_H+m_{AB})^2)}.
    \end{split}
\end{equation}
For the denominators, we can make use of:
\begin{equation}
    \frac{2(k^2-m_{AB}^2-m_H^2)}{(k^2-(m_H-m_{AB})^2)(k^2-(m_H+m_{AB})^2)}=\frac{1}{(k^2-(m_H-m_{AB})^2)}+\frac{1}{(k^2-(m_H+m_{AB})^2)}
\end{equation}
and 
\begin{equation}
    \frac{4m_{AB}m_H}{(k^2-(m_H-m_{AB})^2)(k^2-(m_H+m_{AB})^2)}=-\frac{1}{(k^2-(m_H-m_{AB})^2)}+\frac{1}{(k^2-(m_H+m_{AB})^2)}
\end{equation}.

First, we  calculate the trace 
\begin{multline}
   O=\text{Tr}(\gamma^\mu\left(\gamma^\rho k_{+,\rho} (k_+^2-M^2)-m_{AB}(k_+^2-M^2+2m_H^2)\gamma^3+m_H(k_+^2-M^2+2m_{AB}^2)\gamma^3\gamma^5-2m_{AB}m_H\gamma^\rho k_{+,\rho}\gamma^5\right) \\ \gamma^\nu \left(\gamma^\eta k_{-,\eta} (k_-^2-M^2)-m_{AB}(k_-^2-M^2+2m_H^2)\gamma^3+m_H(k_-^2-M^2+2m_{AB}^2)\gamma^3\gamma^5-2m_{AB}m_H\gamma^\eta k_{-,\eta}\gamma^5\right))
   \label{Eq:OAAtracepart}
\end{multline}
where $k_\pm=k\pm p/2$. First, we write it as 
\begin{equation}
   O=\text{Tr}\left(\gamma^\mu\left(\gamma^\rho k_{+,\rho} A_++B_+\gamma^3+C_+\gamma^3\gamma^5+D_+\gamma^\rho k_{+,\rho}\gamma^5\right)  \gamma^\nu \left(\gamma^\eta k_{-,\eta} A_-+B_-\gamma^3+C_-\gamma^3\gamma^5+D_-\gamma^\eta k_{-,\eta}\gamma^5\right)\right)
\end{equation}
where
\begin{eqnarray}
    A_\pm=&(k_\pm^2-M^2);\\
    B_\pm=&-m_{AB}(k_\pm^2-M^2+2m_{H}^2)\\
    C_\pm=&m_H(k_\pm^2-M^2+2m_{AB}^2)\\
    D_\pm=&-2m_{AB}m_H\\
\end{eqnarray}
 Given that $\text{Tr}(\gamma^\mu\gamma^\nu\gamma^\rho\gamma^3\gamma^5)=-4i\epsilon^{\mu\nu\rho}$, the only non-vanishing terms would have the following form
\begin{equation}
   O=\text{Tr}\left[\gamma^\mu\gamma^\rho\gamma^\nu\gamma^3\gamma^5 k_{+,\rho}A_+C_-+\gamma^\mu\gamma^3\gamma^\nu\gamma^\eta\gamma^5k_{-,\eta}B_+D_-+\gamma^\mu\gamma^3\gamma^5\gamma^\nu\gamma^\eta k_{-,\eta}C_+A_-+\gamma^\mu\gamma^\rho\gamma^5\gamma^\nu\gamma^3k_{+,\rho}D_+B_-\right]
\end{equation}
which can be written as
\begin{equation}
O_{AA}=\text{Tr}\left[\gamma^\mu\gamma^\rho\gamma^\nu\gamma^3\gamma^5\right]\left( k_{+,\rho}A_+C_--k_{-,\rho}B_+D_-- k_{-,\rho}C_+A_-+k_{+,\rho}D_+B_-\right).
\end{equation}
Putting all these terms back in the expression for $\Pi^{\mu\nu}(p)$, we get
\begin{multline}
  \Pi^{\mu\nu}_{AA}(p)= -4i\epsilon^{\mu\rho\nu}\int \frac{d^3k}{(2\pi)^3}\frac{O_{AA}}{(k_+^2-(m_H-m_{AB})^2)(k_+^2-(m_H+m_{AB})^2)(k_-^2-(m_H-m_{AB})^2)(k_-^2-(m_H+m_{AB})^2)}
\end{multline}
First, we consider $k\rightarrow -k$ which transforms $k_\pm\rightarrow -k_\mp$,   $A_\pm\rightarrow A_\mp$, $B_\pm\rightarrow B_\mp$, $C_\pm\rightarrow C_\mp$, and $D_\pm\rightarrow D_\mp$. Next, using the fact that the denominator remains invariant for this transformation we notice that 
\begin{multline}
-k_{-,\rho}B_+D_++k_{+,\rho}B_-D_+=-2m_{AB}m_H\left(\frac{p^\rho}{2}(B_++B_-)+\frac{k}{2}(B_--B_+)\right)\\
=2m_{AB}m_H\left(m_{AB}\frac{p^\rho}{2}(k_+^2+k_-^2-2M^2+4m_H^2)+m_{AB}\frac{k}{2}(k_-^2-k_+^2)\right)
\end{multline}
and
\begin{multline}
k_{+,\rho}A_+C_--k_{-,\rho}C_+A_-=\frac{k^\rho}{2}\left(A_+C_--C_+A_-\right)+\frac{p^\rho}{2}\left(A_+C_-+C_+A_-\right)\\=p^\rho m_H\left(\left[(k_+^2-M^2)(k_-^2-M^2)+m_{AB}^2\left(k_+^2+k_-^2-2M^2\right)\right]\right)-m_{AB}^2m_H k^\rho\left(k_+^2-k_-^2\right)
\end{multline}
\begin{equation}
\implies O_{AA}=p^\rho m_H\left(\left[(k_+^2-M^2)(k_-^2-M^2)+2m_{AB}^2\left(k_+^2+k_-^2-2M^2\right)+4m_{AB}^2m_H^2\right]\right)
\label{OAAexpression}
\end{equation}
which is proportional to the Haldane mass term as expected. 
This results in
\begin{equation}
\begin{split}
 \Pi^{\mu\nu}_{AA}(p) = -i\epsilon^{\mu\rho\nu}m_H p^\rho\int \frac{d^3k}{(2\pi)^3}\,2\left(\frac{1}{k_+^2-m_1^2}\frac{1}{k_-^2-m_1^2}+\frac{1}{k_+^2-m_2^2}\frac{1}{k_-^2-m_2^2}\right)\\-i\epsilon^{\mu\rho\nu}m_{AB} p^\rho\int \frac{d^3k}{(2\pi)^3}\,2\left(\frac{1}{k_+^2-m_2^2}\frac{1}{k_-^2-m_2^2}-\frac{1}{k_+^2-m_1^2}\frac{1}{k_-^2-m_1^2}\right)\\
  = -i\epsilon^{\mu\rho\nu} p^\rho\int \frac{d^3k}{(2\pi)^3}\, 2\left(\frac{m_H+m_{AB}}{(k_+^2-m_2^2)(k_-^2-m_2^2)}+\frac{m_H-m_{AB}}{(k_+^2-m_1^2)(k_-^2-m_1^2)}\right).
\end{split}\end{equation}


\subsection{Phonon-Phonon terms}
In this part, we calculate phonon-phonon correlations:
\begin{equation}
    \Pi^{\mu\nu}_{aa}(p)=\text{Tr}\left(\gamma^\mu\gamma^5\frac{1}{\gamma^\rho(k+p)_\rho-m_D-m_{AB}\gamma^3-m_H\gamma^3\gamma^5}\gamma^\nu\gamma^5\frac{1}{\gamma^\rho k_\rho-m_D-m_{AB}\gamma^3-m_H\gamma^3\gamma^5}\right)
\end{equation}
and we can now borrow most of the results from previous part. Here, we need to calculate the 
\begin{equation}
   O_{aa}=\text{Tr}\left(\gamma^\mu\gamma^5\left(\gamma^\rho k_{+,\rho} A_++B_+\gamma^3+C_+\gamma^3\gamma^5+D_+\gamma^\rho k_{+,\rho}\gamma^5\right)  \gamma^\nu \gamma^5\left(\gamma^\eta k_{-,\eta} A_-+B_-\gamma^3+C_-\gamma^3\gamma^5+D_-\gamma^\eta k_{-,\eta}\gamma^5\right)\right)
\end{equation}
and by using $\{\gamma^\mu,\gamma^5\}=0$ and $(\gamma^5)^2=1$, it becomes
\begin{equation}
   O_{aa}=\text{Tr}\left(\gamma^\mu\left(\gamma^\rho k_{+,\rho} A_++B_+\gamma^3+C_+\gamma^3\gamma^5+D_+\gamma^\rho k_{+,\rho}\gamma^5\right)  \gamma^\nu\left(\gamma^\eta k_{-,\eta} A_-+B_-\gamma^3+C_-\gamma^3\gamma^5+D_-\gamma^\eta k_{-,\eta}\gamma^5\right)\right)=O_{AA}
\end{equation}

\subsection{Phonon-photon terms}
In this section, we calculate cross-correlations between photons and phonons:
\begin{equation}
    \Pi^{\mu\nu}_{aA}(p)=\text{Tr}\left(\gamma^\mu\gamma^5\frac{1}{\gamma^\rho(k+p)_\rho-m_D-m_{AB}\gamma^3-m_H\gamma^3\gamma^5}\gamma^\nu\frac{1}{\gamma^\rho k_\rho-m_D-m_{AB}\gamma^3-m_H\gamma^3\gamma^5}\right).
\end{equation}
Next, we have
\begin{equation}
O_{aA}=\text{Tr}\left[\gamma^\mu\gamma^\rho\gamma^\nu\gamma^3\gamma^5\right]\left(-k_{+,\rho}A_+B_-+ k_{-,\rho}B_+A_-+ k_{-,\rho}C_+D_--k_{+,\rho}D_+C-\right)
\end{equation}
\begin{equation}
 O_{aA}=-4i\epsilon^{\mu\rho\nu}\left(-\frac{p^\rho}{2}\left(B_+A_-+A_+B_-+C_+D_-+D_+C_-\right)+\frac{k}{2}\left(B_+A_--A_+B_-+C_+D_-D+C_-\right)\right)
\end{equation}
\begin{equation}
 O_{aA}=-4i\epsilon^{\mu\rho\nu}\,p^\rho\left(-m_{AB}(k_+^2-M^2)(k_-^2-M^2)-2m_{AB}m_H^2(k_-^2+k_+^2-2M^2)+4m_{AB}{m_{AB}^2}m_H^2\right)
\end{equation}
which is proportional to Semenoff mass term as expected. Interestingly, this expression is similar to Eq.~\ref{OAAexpression} if we swap $m_H\leftrightarrow m_{AB}$. We can directly use that result to show
\begin{equation}
 \Pi^{\mu\nu}_{aA}(p) = -i\epsilon^{\mu\rho\nu} p^\rho\int \frac{d^3k}{(2\pi)^3}\, 2\left(\frac{m_H+m_{AB}}{(k_+^2-m_1^2)(k_-^2-m_1^2)}+\frac{m_{AB}-m_{H}}{(k_+^2-m_1^2)(k_-^2-m_1^2)}\right).
\end{equation}
Using the properties of gamma matrices, we can show 
\begin{equation}
    \Pi^{\mu\nu}_{Aa}(p)=\Pi^{\nu\mu}_{aA}(-p).
\end{equation}
\section{Equation for splitting}
In this section, we derive the splitting of chiral phonons using the equation of motion:
\begin{eqnarray}
    \ddot{a}_{L/R}+\omega_0^2a_{L/R}=\pm 2 i \frac{g_a^2}{2\pi}\left[c\dot{a}_{L/R}+d\dddot{a}_{L/R}\right]\\\implies
    -\omega^2+\omega_0^2a_{L/R}=\pm 2  \frac{g_a^2}{2\pi}\left[-c\omega+d\omega\right]
\end{eqnarray}
and the two chiral modes satisfy:
\begin{eqnarray}
    \omega_L^2-\omega_0^2= 2\alpha  \left[c\omega_L-d\omega_L^3\right]\\
     \omega_R^2-\omega_0^2= -2\alpha  \left[c\omega_R-d\omega_R^3\right]
\end{eqnarray}
where $\alpha=g_a^2/(2\pi)$. We assume a solution of the form $\omega_{L/R}=\omega_A\pm \eta$
\begin{eqnarray}
    \omega_A^2+2\omega_A\eta+\eta^2-\omega_0^2= 2\alpha  \left[c\omega_0+c\eta-d\omega_0^3-3d\omega_0^2\eta-3d\omega_0\eta^2-\eta^3d\right]\\
    \omega_A^2-2\omega_A\eta+\eta^2-\omega_0^2= -2\alpha  \left[c\omega_0-c\eta-d\omega_0^3+3d\omega_0^2\eta-3d\omega_0\eta^2+\eta^3d\right]
\end{eqnarray}
which gives:
\begin{eqnarray}
    2\omega_A\eta= 2\alpha  \left[c\omega_0-d\omega_0^3-3d\omega_0\eta^2\right]
\end{eqnarray}
and for small splitting, $\eta\approx \alpha\left[c-d\omega_0^2\right]$. 

If we take into account the spatial variations in the second term with $p^2=\partial_t^2-v_F^2\partial_x^2$,
\begin{equation}
    \ddot{a}_{L/R}+\omega_0^2a_{L/R}=\pm 2 i \frac{g_a^2}{2\pi}\left[c\dot{a}_{L/R}+d(\dddot{a}_{L/R}-v_F^2 \nabla^2 \dot{a}_{L/R}\right]
\end{equation}
which can be expressed as
\begin{eqnarray}
    \omega_L^2-\omega_0^2= 2\alpha  \left[c\omega_L-d\omega_L^3+d\omega_L v_F^2 k^2\right]\\
     \omega_R^2-\omega_0^2= -2\alpha  \left[c\omega_R-d\omega_R^3+d\omega_R v_F^2 k^2\right]
\end{eqnarray}
in the Fourier space.  As a result, the modes become weakly dispersive with different group velocities for two chiralities.

\section{Equation of motion}
\label{appendix:EOM}
The full effective Lagrangian density for EM and phonon fields
\begin{equation}
\begin{split}
    \mathcal{L}_{eff}(A,a)&=\frac{1}{2}\frac{e^2v_F^2}{g^2}\sum_{m=x,y}\left(\rho_I(\partial_t a_m)^2-\rho_I\omega_0^2a_m^2\right)-\frac{1}{4}F_{\mu\nu}F^{\mu\nu}\\&-\frac{e^2}{2\pi\hbar}\sum_{i=1,2}\left( c(m_i) (A_x\dot{A}_y-A_y\dot{A}_x)+d(m_i) (A_x\dddot{A}_y-A_y\dddot{A}_x)-v_F^2d(m_i) (A_x\nabla^2\dot{A}_y-A_y\nabla^2\dot{A}_x)\right)\\&+\frac{e^2}{2\pi\hbar}\sum_{i=1,2} c(m_i) (a_x\dot{a}_y-a_y\dot{a}_x)+d(m_i) (a_x\dddot{a}_y-a_y\dddot{a}_x)-v_F^2d(m_i) (a_x\nabla^2\dot{a}_y-a_y\nabla^2\dot{a}_x)\\&+\frac{e^2}{2\pi\hbar}\sum_{i} \left(c(\tilde{m}_i) (a_x\dot{A}_y-A_y\dot{a}_x+A_x\dot{a}_y-a_y\dot{A}_x)\right)+\frac{e^2}{2\pi\hbar}\sum_{i=1,2}d(\tilde{m}_i) (a_x\dddot{A}_y-A_y\dddot{a}_x+A_x\dddot{a}_y-a_y\dddot{A}_x)\\&-\frac{e^2}{2\pi\hbar}\sum_{i=1,2}v_F^2d(\tilde{m}_i) (a_x\nabla^2\dot{A}_y-A_y\nabla^2\dot{a}_x+A_x\nabla^2\dot{a}_y-a_y\nabla^2\dot{A}_x)
\end{split}
\end{equation}
where $m_{1/2}=m_H\pm m_{AB}$ and $\tilde{m}_i=m_{AB}\pm m_H$. The equation of motion can be derived using higher-order Euler-Lagrangian equations:
\begin{equation}
\frac{\partial \mathcal{L}}{\partial a_i}
- \frac{\partial}{\partial t} \left( \frac{\partial \mathcal{L}}{\partial \dot{a}_i} \right)
+ \frac{d^2}{dt^2} \left( \frac{\partial \mathcal{L}}{\partial \ddot{a}_i} \right)
- \frac{\partial^3}{\partial t^3} \left( \frac{\partial \mathcal{L}}{\partial \dddot{a}_i} \right)-\frac{\partial^3}{\partial t\partial x^2} \left( \frac{\partial \mathcal{L}}{\partial (\partial_x^2\dot{a}_i)} \right)-\frac{\partial^3}{\partial t\partial y^2} \left( \frac{\partial \mathcal{L}}{\partial (\partial_y^2\dot{a}_i)} \right)
= 0, \qquad i = x, y
\end{equation}
which leads to 
\begin{eqnarray}
\rho_I \left( \ddot{a}_x + \omega_0^2 a_x \right) = \frac{2g^2}{hv_F^2} C \dot{a}_y + \frac{2g^2}{hv_F^2} D (\dddot{a}_y-v_F^2\nabla^2\dot{a}_y)+  \frac{2g^2}{hv_F^2}\tilde{C}\dot{A}_y+ \frac{2g^2}{hv_F^2}\tilde{D}(\dddot{A}_y-v_F^2\nabla^2\dot{A}_y),\\
\rho_I \left( \ddot{a}_y + \omega_0^2 a_y \right) = - \frac{2g^2}{hv_F^2}  C \dot{a}_x- \frac{2g^2}{hv_F^2}  D (\dddot{a}_x-v_F^2\nabla^2\dot{a}_x) - \frac{2g^2}{hv_F^2}\tilde{C}\dot{A}_x- \frac{2g^2}{hv_F^2}\tilde{D}(\dddot{A}_x-v_F^2\nabla^2\dot{A}_x),
\end{eqnarray}
In chiral basis: 
\begin{eqnarray}
\begin{split}
\rho_I \left( \ddot{a}_{L/R} + \omega_0^2 a_{L/R} \right) = \mp i\,\frac{2g^2}{hv_F^2}\left(C \dot{a}_{L/R} +D(\dddot{a}_{L/R} -v_F^2\nabla^2\dot{a}_{L/R})\right) \mp i\,\frac{2g^2}{hv_F^2}\left(\tilde{C}\dot{A}_{L/R}+\tilde{D}(\dddot{A}_{L/R}-v_F^2\nabla^2\dot{A}_{L/R})\right)
\end{split}
\end{eqnarray}
where $C=\sum_{i=1,2}c(m_i),\,\tilde{C}=\sum_{i=1,2}c(\tilde{m_i}),\,D=\sum_{i=1,2}d(m_i)$ and $\tilde{D}=\sum_{i=1,2}c(\tilde{m}_i)$ which are defined below Eq.~\ref{Eq:Gammamunu}.

\bibliography{ref.bib}
\end{document}